# Correlating Research Octane Numbers of Gasoline Surrogates with High Temperature Oxidation Characteristics

Ruijie Zhu

*Qianweichang College, Shanghai University, Shanghai 200444, China*

Due to the rapid combustion nature of gasolines, most of the experimental measurements of research octane number (RON) could only yield macroscopic quantities such as ignition delay time and pressure variations, but fail to yield microscopic quantities such as the time evolution of number of species, which limits deep understanding of combustion mechanisms. In this work, molecular dynamics simulation was used to unveil the correlation between the RON of gasoline surrogates and their combustion behaviors. Results show that the time of turning point obtained from potential energy profile exhibits a strong linear correlation with RON, and the number of hydroxyl radicals per molecule at equilibrium has a clear negative correlation with RON, making both of them good features for predicting the antiknock ability of fuel surrogates.

## I. Introduction

In spark-ignition engines, engine knocking occurs when end-gas auto-ignites before being reached by the flame front[1]. Knocking leads to numerous negative effects, including reduced engine performance and thermal efficiency[2]. It is a complex phenomenon influenced by many factors, such as the design of combustion chamber and the antiknock ability of gasolines. The latter is typically quantified using motor octane number (MON) and research octane number (RON), both of which are commonly determined by combusting gasolines in a cooperative fuel research (CFR) engine following the ASTM standard[3]. Compared to MON, RON describes gasoline's combustion behavior at a lower engine temperature and speed, making it a better metric for representing the real-world engine operation condition[4]. Herein, RON is chosen as the evaluation metric for fuel's ability to resist engine knocking. Experimental determination of RON is both time-consuming and expansive. Besides, due to the fast combustion nature of gasoline surrogates, conventional experimental analysis is limited to macroscopic quantities such as ignition delay time and pressure variations, whereas a thorough interpretation of microscopic combustion mechanisms is absent[5-7]. To address the drawbacks of experimental methods, kinetic simulation was adopted in this work to model the chemical reactions during combustion, and eventually making predictions for RON[8-11]. The time evolution profiles of species during combustion simulations are crucial to the understanding of microscopic mechanisms. In this work, ReaxFF-based molecular dynamics simulation was adopted to model high temperature combustion reactions of gasoline surrogates. The ReaxFF reactive forcefield developed by Chenoweth et al.[12] has been widely applied to the simulation of high temperature pyrolysis and oxidation behavior of hydrocarbons[13-16], thus it serves as an alternative to kinetic simulations.

Commercial gasolines contain a wide variety of components, including n-paraffins, iso-paraffins, olefins, naphthenes and aromatics. Gasoline surrogates are commonly used to simplify modeling, which contains one or more components. Surrogate models are designed such that they reproduce the combustion behavior of real fuels. One type of surrogate fuel is the primary reference fuel (PRF), which consists of a binary mixture of n-heptane and iso-octane with RON ranging from 0 to 100. PRF serves as a standard for RON determination. Table 1 lists the hydrocarbons used in this study together with their densities at 300K.

Table 1 RON and densities of hydrocarbons[17]

| Hydrocarbon | RON | Density(g/cm$^3$) |
|---|---|---|
| n-butane | 94 | 0.580 |
| n-pentane | 62 | 0.626 |
| n-hexane | 25 | 0.659 |
| n-heptane | 0 | 0.684 |
| n-octane | -20 | 0.703 |
| iso-butane | 102 | 0.551 |
| iso-pentane | 92.3 | 0.620 |
| iso-hexane | 73.4 | 0.653 |
| iso-heptane | 44 | 0.679 |
| iso-octane | 100 | 0.692 |
| 1-butene | 97.4 | 0.577 |
| 1-pentene | 90 | 0.641 |
| 1-hexene | 77 | 0.670 |
| 1-heptene | 60 | 0.697 |
| 1-octene | 28.7 | 0.715 |
| toluene | 118 | 0.866 |
| cyclohexane | 83 | 0.778 |

For n-paraffins, iso-paraffins and olefins, the number of carbon atoms per molecule ranges from 4 to 8, with RON ranging from -20 to 100. Toluene and cyclohexane were chosen as representative molecules for aromatics and nephthene, respectively.

## II. Simulation Details

For each type of hydrocarbon molecule, a simulation box containing 300 molecules was generated using the amorphous builder module of Materials Studio[18]. The simulation density was chosen to be half of the density of pure hydrocarbons at 300 K. This ensures sufficient contact of hydrocarbon molecules with oxygen. The hydrocarbon / oxygen ratio in the simulation box is set to stoichiometry to promote complete combustion reaction. As an example, the simulation box of iso-butane is shown in Fig. 1. To find a proper configuration, geometry optimization was carried out using universal force field with ultrafine accuracy.

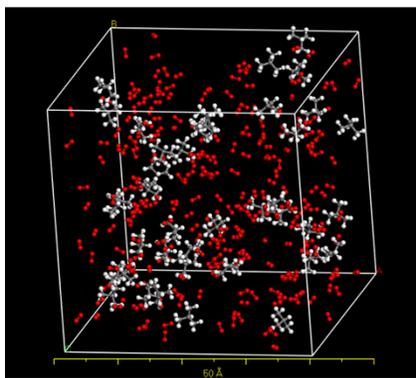

Fig. 1 Simulation box of iso-butane

To mimic high temperature oxidation reaction in the combustion chamber, the heating temperature should neither be too low nor too high. A too low heating temperature would result in insufficient combustion, whereas a too high heating temperature would result in unphysical combustion rate. In this study, 2500 K and 2750 K were chosen as heating temperatures, because both temperatures ensure the occurrence of oxidation reactions with a proper reaction rate. The molecular dynamics simulation consists of four stages: energy minimization, NPT for 100 ps, heating from 300 K to 2500 K / 2750 K (NVT) within 10 ps, and NVE for 1000 ps. During energy minimization, conjugate gradient algorithm was used along with force and energy convergence criteria of $10^{-6}$ kcal/(mol·Å) and $10^{-6}$ kcal/mol, respectively. For 100 ps NPT, pressure and temperature were set to standard atmospheric pressure (1bar) and 300 K. For 10 ps NVT heating and 1000 ps NVE, the timestep is 0.1 fs, and results are outputted every 1000 steps.

### 2.1 High temperature oxidation simulation of hydrocarbons with heating temperature of 2500K

Potential energy is a good indicator of the combustion state of simulated system. Fig. 2 shows the potential energy profile of iso-butane during NVE simulation after being heated to 2500 K. Since the number of carbon atoms in the molecule greatly affects RON, the potential energy values are normalized to per carbon atom to ensure fair comparison among different hydrocarbon systems. The initial state (t=0 ps) of the potential energy curve was chosen as the time at the end of heating, with its value set to zero. Ten point moving average was used to smooth out short-term fluctuations. As shown in Fig. 2, each potential energy profile could be subdivided into two parts, including a plateau and a linear decreasing part. For each hydrocarbon system, the time of turning point, which signifies the moment the simulated system undergoes significant changes, was extracted (Fig. 3).

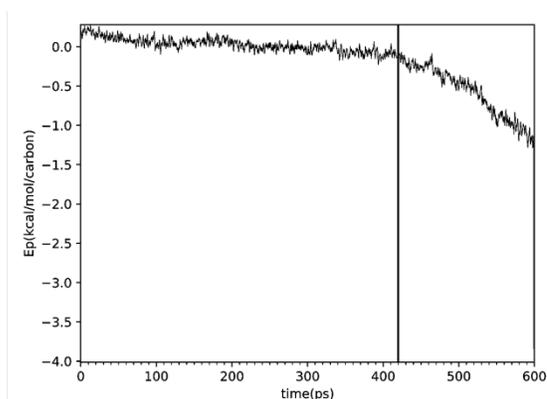

Fig. 2 Potential energy profile of iso-butane during NVE simulation

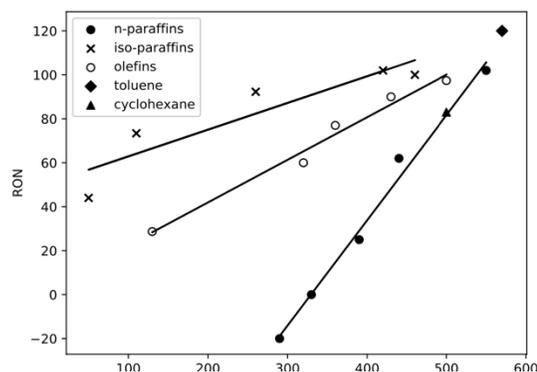

Fig. 3 Relationship between time of turning point and RON. The solid lines are linear fits for the data points.

As shown in Fig. 3, for linear-chain hydrocarbons (n-paraffins and olefins), a strong positive linear relationship is observed between time of turning point and RON, with a linear correlation coefficient of 0.99 and 0.98, respectively. For iso-paraffins, a type of branched hydrocarbons, monotonic relationship between the two quantities is observed, with a linear correlation coefficient of 0.83. It is worth noting that cyclohexane also obeys the linear trend for n-paraffins, which might be due to its transformation to linear chains upon breaking C-C bonds. Besides, the RON for branched hydrocarbons are always higher than their linear counterparts, which supports the recent finding that isomerization may improve the antiknock ability of gasolines[19].

During high temperature oxidation of hydrocarbons, a series of intermediate species were generated, such as fragments of hydrocarbons, hydroxyl radicals and hydroperoxyl radicals. However, for most hydrocarbon systems, NVE simulation with a heating temperature of 2500 K could only yield a small number of intermediate species, which limits the feasibility of fragment analysis. Thus, a higher heating temperature is necessary to reveal the underlying relationship between RON and time evolution of the number of species. In this regard, 2750K was chosen as the heating temperature for the next set of oxidation simulations.

## 2.2 High temperature oxidation simulation of hydrocarbons with heating temperature of 2750K

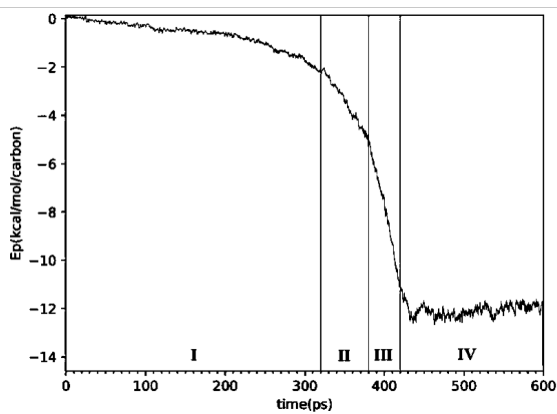

Fig. 4 Potential energy profile of iso-butane during NVE simulation

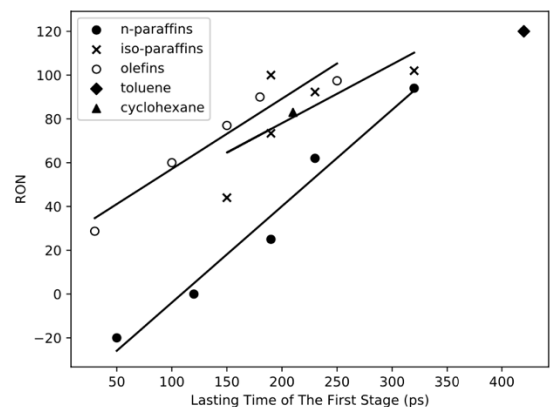

Fig. 5 Relationship between lasting time of the first stage and RON. The solid lines are linear fits to data points.

The increase of heating temperature from 2500 K to 2750 K results in a faster combustion rate, which accounts for the decrease of the lasting time for the stage I, as shown in Fig. 4. Compared with stage I, stage II to IV are less important because they reveal the change of the simulated systems during combustion, rather than resistance to combustion. For n-paraffins and olefins, the monotonic trend between the lasting time of stage I and RON is still preserved (Fig. 5), whereas the positive linear relationship is a bit weaker, with a correlation coefficient of 0.97 and 0.94, respectively.

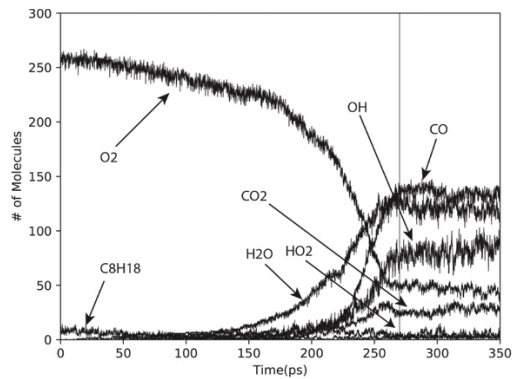

Fig. 6 Time evolution profile of species for iso-octane during NVE simulation

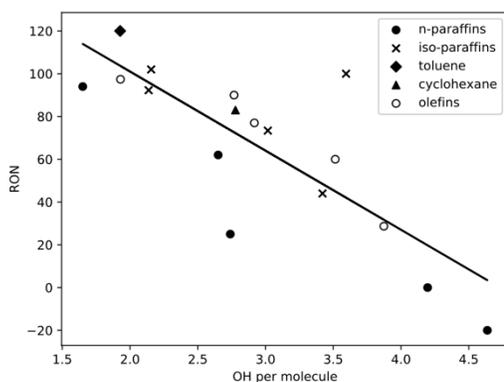

Fig. 7 Correlation between the equilibrium number of hydroxyl radials per molecule and RON. The solid line is linear fit to data points.

The time evolution profile of species for iso-octane during NVE simulation is shown in Fig. 6. A large number of intermediate species were generated during combustion, including fragments of hydrocarbons, ·OH radicals, HO$_2$· radicals as well as molecules such as H$_2$O, CO, and CO$_2$. Although both radicals were known to be critical for oxidation reactions, the number of ·OH per molecules at equilibrium was observed to be much larger than that of HO$_2$· at 2750 K, making it less prone to be affected by thermodynamic fluctuation. Therefore, the number ·OH per molecules at equilibrium was used to correlate to RON, as shown in Fig. 7. For all hydrocarbons systems, a negative correlation between the number of ·OH per molecules at equilibrium and RON is observed, with a higher value corresponding to a lower RON. It is worth noting that such an inverse relationship between the two quantities applies to not only linear-chain hydrocarbons but also branched-chain hydrocarbons, making it a general feature for the task.

## III. Conclusions

In this work, high temperature molecular dynamics simulations with heating temperatures of 2500 K and 2750 K were carried out to reveal the correlation between RON of gasoline surrogates and their combustion characteristics. Two features obtained from simulations were identified to have a clear correlation with RON. The first feature, time of turning point from the potential energy profile, was found to have a positive linear correlation with RON. It signifies the lasting time of the simulated system before going through significant changes. The other feature is the number of hydroxyl radicals per molecule in equilibrium, as obtained from the time evolution profile of species, where a clear negative correlation with RON is observed. Since both features could be determined directly from molecular dynamics simulations without performing experiments, they could potentially be used for pre-assessment of the antiknock ability of fuels.